\documentclass[reprint,prd,nofootinbib,preprintnumbers]{revtex4}
 \usepackage{graphicx,psfrag,epsfig,epsf,colordvi,slashed,amsmath,bm}

 \def\lsim{\mathrel{\vcenter{\hbox{$<$}\nointerlineskip\hbox{$\sim$}}}}
\def\gsim{\mathrel{\vcenter{\hbox{$>$}\nointerlineskip\hbox{$\sim$}}}}

\begin{document}
\preprint{MAN/HEP/2012/020}
\title{{\Large Light and Superlight Sterile Neutrinos}\\[2mm]
{\Large in the Minimal Radiative Inverse Seesaw Model}
\vspace{0.3cm}}

\author {\bf  P. S. Bhupal Dev and Apostolos Pilaftsis\\}
%\vspace{0.3in}
%
\affiliation{\vspace{2mm}
Consortium for Fundamental Physics, School of Physics and Astronomy,
University of Manchester, Manchester, M13 9PL, United Kingdom} 

\begin{abstract}
\vspace{0.2cm}
\centerline{\bf ABSTRACT}
\vspace{1mm}
\noindent
We explore  the possibility of light and  superlight sterile neutrinos
in the recently proposed Minimal Radiative Inverse Seesaw extension of
the Standard  Model for neutrino  masses, in which {\it  all} existing
neutrino data  can be explained.  In particular, we  discuss two benchmark
scenarios with one of the three sterile neutrino states  in  the 
keV-range,  having very
small mixing with the active  neutrinos to account for the Dark Matter
in the Universe, while (i) the other two light sterile  neutrino states 
could be in the
eV-range,  possessing  a nonzero  mixing  with  the  active states  as
required  to  explain the  LSND+MiniBooNE+reactor  neutrino data, or   
(ii) one of the light sterile states is in the eV-range, whereas the second 
one could be superlight and almost mass-degenerate with the
solar neutrinos. Such superlight sterile neutrinos could give rise to
potentially   observable  effects   in  future   neutrino  oscillation
experiments and  may also offer  a possible explanation for  the extra
radiation observed in the Universe. 
\end{abstract}

\maketitle

\section{Introduction}

A large number of solar, atmospheric, reactor and accelerator neutrino
experiments  provide  solid  pieces   of  evidence  for  the  neutrino
oscillation  phenomenon, and  hence for  nonzero neutrino  masses and
mixing~\cite{pdg}.   Most  of   these  experimental  results  can  be
understood  by  oscillations  of  the  three  so-called  {\em  active}
neutrinos,   which  appear  in   charged-  and   neutral-current  weak
interactions    in    the    Standard    Model   (SM).     A    global
fit~\cite{global,othergf} to the three-neutrino oscillation data gives
their  best-fit  mass-squared differences  of  $\Delta m^2_{\rm  sol}=
7.5\times  10^{-5}~{\rm eV}^2$  and $\Delta  m^2_{\rm  atm}= 2.4\times
10^{-3}~{\rm eV}^2$.   However, there are a  few experimental results,
most  notably  the  LSND  anomaly~\cite{lsnd} and  more  recently  the
MiniBooNE    results~\cite{mini},    as    well   as    the    reactor anti-neutrino 
anomaly~\cite{reactor},   which  cannot   be  explained   within  this
three-neutrino-mixing   paradigm  and   seem  to   require  additional
neutrino(s)  with mass-squared differences  of order 1 eV$^2$.
Such additional neutrinos cannot  couple directly to the SM $Z$-boson,
due  to the  LEP constraint  on its  invisible  decay width~\cite{lep}
which allows only three active light neutrinos. Hence, these neutrinos
must be {\em  sterile} neutrinos~\cite{white}, i.e.~SM gauge singlets,
and can  only participate in  the SM weak  interactions through their
mixing with the active neutrinos.

In 
the  light  of  recent   predictions  of  a  slightly  higher  reactor
anti-neutrino  flux~\cite{mueller}, a global analysis of the short baseline (SBL) neutrino oscillation data  favours the  existence  of  more  than one  light
sterile  neutrino in  the eV  mass range~\cite{kopp,  giunti, 
  donini,  conrad, fornengo}.   On the  other hand,  cosmological data
also  indicate  a weak  preference  for  additional  light degrees  of
freedom:  $N_{\rm   eff}=4.34\pm  0.87$~\cite{wmap},  which   could  be
interpreted as another evidence for sterile neutrinos in the eV/sub-eV
mass range~\cite{hamann,fornengo}.
	  
The existence of light  sterile neutrinos requires further theoretical
justification,  since  unlike   the  active  neutrinos,  the  sterile
neutrino masses are not protected by the SM gauge symmetry.  There are
a      number      of      interesting      proposals      in      the
literature~\cite{white}\footnote{Light  sterile   neutrinos  can  also
  occur  in   theories  of   quantum  gravity  due   to  gravitational
  interactions   involving  global   anomalies   related  to   quantum
  torsion~\cite{MP}.}  to   explain  the  lightness   of  the  sterile
neutrinos  in  various  extensions  of  the SM,  for  nonzero  active
neutrino  masses $\nu_{iL}$  (with $i=e,\mu,\tau$).   In  the simplest
extension  which realizes  the  type-I seesaw  mechanism~\cite{type1},
$n_R$     right-handed     neutrinos     $\nu_{\alpha    R}$     (with
$\alpha=1,2,...,n_R$)  are added  to the  SM  some of  which could  in
principle  play the role  of light  sterile neutrinos,  provided their
masses are  in the eV range~\cite{degouvea05}.  The  masses and mixing
of  the  active  and  sterile   neutrinos  can  be  deduced  from  the $(3+n_R)\times (3+n_R)$  complex symmetric neutrino mass matrix  in the flavour
basis $\{(\nu_{iL})^C,\nu_{\alpha R}\}$:
\begin{eqnarray}
{\cal M}_\nu\ =\ \left(\begin{array}{cc}
{\bf 0} & M_D \\
M_D^{\sf T} & M_R
\end{array}\right)\; ,
\label{eq:Mnutype1}
\end{eqnarray}
where  $M_D$  is the  Dirac  neutrino mass  matrix,  and  $M_R$ is  the
$(B-L)$-breaking Majorana mass matrix of the singlet neutrinos. In the
usual  seesaw approximation  $\|\xi\| \equiv  \|M_DM_R^{-1}\|  \ll 1$,
with $\|\xi\|\equiv  \sqrt{{\rm Tr}(\xi^\dag \xi)}$ being  the norm of
the matrix  $\xi$, the light  neutrino masses are determined  from the
mass matrix
\begin{eqnarray}
M_{\nu_L}\ \simeq\ -\,M_D M_R^{-1} M_D^{\sf T}\; ,
\label{eq:type1}
\end{eqnarray}
with  mass eigenvalues  $m_{1,2,3}$,  whereas the  heavier masses  are roughly  the  eigenvalues of $M_R$: $m_{4,5,...,3+n_R}\gg  m_{1,2,3}$. In
this  case,   the  active-sterile   mixing,  given  by   $\Theta  \sim
(m_{1,2,3}/m_{4,5,...,n_R})^{1/2}$, is  very small, thus  limiting our
ability  to observe  the singlet  neutrinos, unless  $M_R\lsim  10$ eV
which  are  mostly   excluded  over  a  wide  range   by  the  present
data~\cite{degouvea09}.

It  is  therefore  interesting  to  examine  a  theoretical  framework
yielding  simultaneously light  sterile neutrinos  as favoured  by the SBL
neutrino oscillation  data and a  sizable active-sterile mixing  to be
observable in  the current or near future  experiments.  An attractive
possibility  is  the  so-called inverse  seesaw  model~\cite{inverse},
where  in addition  to the  right-handed  neutrinos as  in the  type-I
seesaw model,  another set of  SM singlet fermions $S_{\rho  L}$ (with
$\rho=1,2,...,n_S$)  is  introduced. In  this  case,  the mass  matrix
in~(\ref{eq:Mnutype1})     gets    extended    to     the    following
$(3+n_R+n_S)\times   (3+n_R+n_S)$   general mass   matrix   in    the   basis
$\{(\nu_{iL})^C,\nu_{\alpha R},(S_{\rho L})^C \}$:
\begin{eqnarray}
{\cal M}_\nu\ =\ \left(\begin{array}{ccc}
{\bf 0} & M_D & {\bf 0}\\
M_D^{\sf T} & \mu_R &  M_N^{\sf T}\\
{\bf 0} & M_N & \mu_S 
\end{array}\right)\; ,
\label{eq:inverse1}
\end{eqnarray}
where  in  addition to  the  usual Dirac  mass  matrix  $M_D$ and  the
Majorana mass  matrix $\mu_R$ in the $\nu_L$--$\nu_R$  sector, we have
added another Dirac mass matrix $M_N$ and Majorana mass matrix $\mu_S$
in the $\nu_R$--$S$ singlet sector.  Observe that the standard inverse
seesaw model discussed originally in~\cite{inverse} is recovered, once
we  set the  right-handed neutrino  Majorana  mass $\mu_R  = {\bf  0}$
in~(\ref{eq:inverse1}).   In this  case, for  $\|\mu_S\|  \ll \|M_N\|,
\|M_D\|$, it  is possible to  have very light sterile  neutrinos, e.g.
in  the  presence  of  a  $\mu-\tau$  symmetry~\cite{rabi}  or  in
theories with warped extra dimensions~\cite{sung}.

Another interesting realization of  inverse seesaw models arises, when
$\mu_R\neq {\bf 0}$,  but $\mu_S={\bf 0}$~\footnote{This structure may
  arise, for  instance, in  models where the  Majorana masses  for the
  $S$-fields    are     forbidden    due    to     specific    flavour
  symmetries~\cite{chun}.}. In this case,  the rank of the mass matrix
given  by~(\ref{eq:inverse1})  reduces   to  $3+n_R$,  and  the  light
neutrinos are {\it exactly} massless  at the tree level. However, they
acquire a  small mass  at the one-loop  level which is  {\em directly}
proportional  to  the Majorana  mass  matrix $\mu_R$~\cite{model}  and
results from  well-known SM  radiative corrections involving  the $Z$-
and  Higgs  bosons~\cite{ap92}.   This  scenario, called  the  Minimal
Radiative  Inverse  Seesaw  Model  (MRISM)  in~\cite{model},  is  very
economical, as it does not require the existence of other non-standard
scalar or gauge fields or other fermionic matter beyond  the singlet
neutrinos~$\{\nu_{\alpha R},S_{\rho L}\}$.

In  this paper,  we  study  the possibility  of  having light  sterile
neutrinos  in  the  MRISM   with  three  pairs  of  singlet  neutrinos
$\{\nu_{iL},S_{\rho   L}\}$  (for   $i,\rho=1,2,3$).   In   the  limit
$\|\mu_R\|\gg \|M_D\|,\|M_N\|$, after  integrating out the three heavy
singlet states $\nu_{1,2,3R}$ with masses of order $\|\mu_R\|$, we are
left with  six light  mass states.  Of  these six light  states, three
should describe  the mostly active  neutrinos and the  remaining three
will  be  mostly  sterile  states.   In  particular,  we  discuss  two
benchmark scenarios.  In the first scenario, all the sterile neutrinos
are heavier  than the active ones,  with two of them having mass 
in the eV-range (as in the 3+2 models) to explain the LSND+MiniBooNE+reactor 
data, while the remaining one is in the  keV-range to  account for  the  Dark   
Matter  (DM)   in  the
Universe~\cite{Asaka:2006ek}. In the second scenario, 
one of the sterile  neutrinos is in the  keV-range as in  the first scenario, 
and the second one is in the eV-range (as in the 3+1 models), while  the third sterile  neutrino is {\it superlight} and  almost mass-degenerate
with the electron neutrino $\nu_{eL}$.  We present numerical estimates
of the model parameters for both of these benchmark scenarios. Finally, we also comment on another possibility discussed in the literature when all the three sterile states are in the eV-range (the 3+3 models).  

The paper is organized  as follows. In Section~\ref{model}, we briefly
review the MRISM and present  the effective neutrino mass matrix after
integrating out the heavy  sterile states. In Section~\ref{simple}, we
analyze  the  simple case  of  one  single  flavour, for  illustration
purposes. In Section~\ref{general}, we  generalize our analysis to the
three  flavour case and  discuss the  experimental constraints  on the
mixing parameters. In  Section~\ref{benchmark}, we present our results
for the benchmark  scenarios mentioned above.  Our conclusions are
given  in Section~\ref{conclude}.  In  Appendix A,  we give  a general
parametrization of the unitary matrix used in our analysis.

\section{Sterile Neutrinos in the MRISM}\label{model}

In  the  limit   $\|\mu_R\|\gg  \|M_D\|,\|M_N\|$,  the  $\nu_R$-fields
decouple  below the  mass  scale $\mu_R$,  resulting  in an  effective
theory   with   six   neutrino   states:  $\nu_{e,\mu,\tau   L}$   and
$S_{1,2,3L}$.  At  the tree level, the effective  neutrino mass matrix
in the weak basis $\{ \nu_{e,\mu,\tau L},S_{1,2,3 L}\}$ becomes
\begin{eqnarray}
{\cal M}_{\rm eff}^{\rm tree} = \left(\begin{array}{cc} 
M_D \mu_R^{-1} M_D^{\sf T} & M_D \mu_R^{-1}M_N^{\sf T} \\
M_N\mu_R^{-1}M_D^{\sf T} & M_N\mu_R^{-1}M_N^{\sf T}
\end{array}\right)
 = \left(\begin{array}{c} 
M_D \\ M_N \end{array}\right) \mu_R^{-1} \left(\begin{array}{cc} 
M_D^{\sf T} & M_N^{\sf T}\end{array}\right)\; . 
\label{eq:Meff}
\end{eqnarray}
Note  that  one  of  the  block eigenvalues  must  vanish,  since  the
effective mass matrix ${\cal M}_{\rm eff}^{\rm tree}$ has rank 3.  The
matrix ${\cal M}_{\rm eff}^{\rm  tree}$ can be block-diagonalized by a
unitary transformation:
\begin{eqnarray}
{\cal V}^{\sf T} {\cal M}_{\rm eff}^{\rm tree} {\cal V} = \left(\begin{array}{cc}
{\bf 0}_3 & {\bf 0}_3 \\
{\bf 0}_3 & M\mu_R^{-1}M^{\sf T} 
\end{array}\right),
\label{eq:block}
\end{eqnarray}
where the  unitary matrix  ${\cal V}$ has  an exact  representation in
terms of an arbitrary matrix $\zeta$~\cite{ap93}:
\begin{eqnarray}
\label{eq:V0}
{\cal V} = \left(\begin{array}{ccc}
({\bf 1}_3+\zeta^*\zeta^{\sf T})^{-1/2} & \zeta^*({\bf 1}_3+\zeta^{\sf T}\zeta^*)^{-1/2}\\
-\zeta^{\sf T}({\bf 1}_3+\zeta^*\zeta^{\sf T})^{-1/2} & ({\bf 1}_3+\zeta^{\sf T}\zeta^*)^{-1/2}
\end{array}\right)\;.
\label{eq:V}
\end{eqnarray} 
From~(\ref{eq:block}), it is straightforward to obtain 
\begin{eqnarray}
\zeta = M_DM_N^{-1}\quad \mbox{and}\quad
M = ({\bf 1}_3+\zeta^\dag \zeta)^{1/2}\,M_N\; .
\end{eqnarray}

\begin{figure}
\centering
\includegraphics[width=7cm]{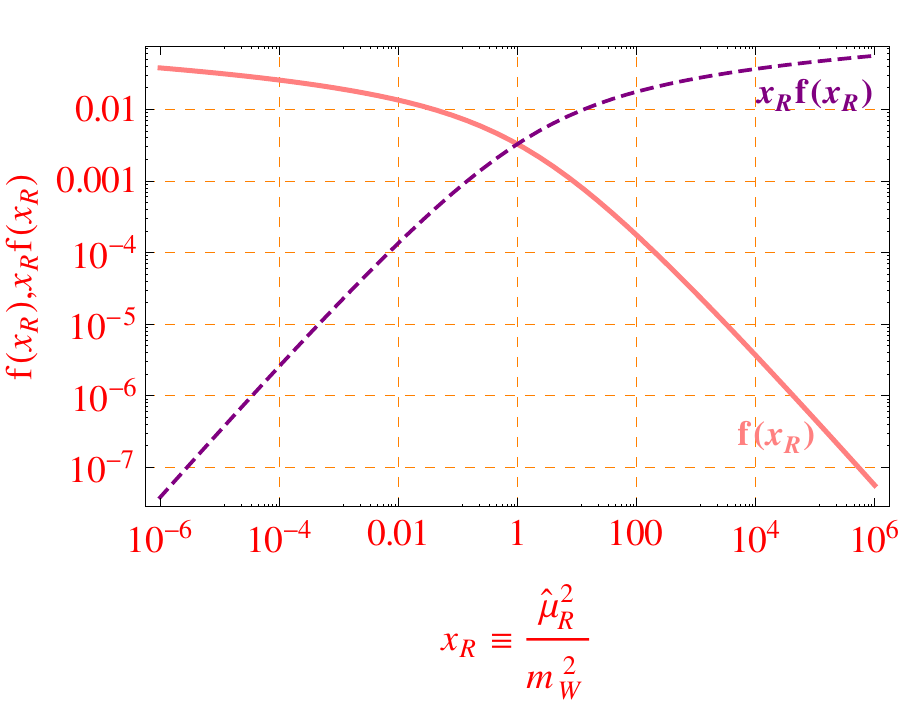}
\caption{The analytic dependence of the one-loop function $f(x_R)$ and
  $x_Rf(x_R)$ on the Majorana mass scale $\hat{\mu}_R$.} 
\label{fig:2a}
\end{figure}

At  the  one-loop  level,  the mass  matrix  given  by~(\ref{eq:Meff})
receives   an  electroweak   radiative   correction  proportional   to
$\mu_R$~\cite{model},    which   in    the   limit    $\|\mu_R\|   \gg
\|M_D\|,\|M_N\|$ is given by
\begin{eqnarray}
{\cal M}_{\rm eff}^{1-{\rm loop}} &\simeq & \left(\begin{array}{c} M_D
  \\ {\bf 0}_3 \end{array}\right)\frac{\alpha_W}{16\pi
  m_W^2}\mu_R\left[\frac{m_H^2}{\mu_R^2-m_H^2{\bf 1}_3}\ln
  \left(\frac{\mu_R^2}{m_H^2}\right) +   
	\frac{3m_Z^2}{ \mu_R^2-m_Z^2{\bf 1}_3}\ln\left(\frac{
          \mu_R^2}{m_Z^2}\right)\right]\left(\begin{array}{cc}
  M_D^{\sf T} & {\bf 0}_3\end{array}\right)\nonumber\\ 
& = & 
\left(\begin{array}{cc} 
M_D \mu_R^{-1} x_Rf(x_R) M_D^{\sf T} & {\bf 0}_3 \\
{\bf 0}_3 & {\bf 0}_3
\end{array}\right) \equiv \left(\begin{array}{cc} 
\Delta M & {\bf 0}_3 \\
{\bf 0}_3 & {\bf 0}_3
\end{array}\right) 
\; ,
\end{eqnarray}
where the one-loop function $f(x_R)$ is defined as 
\begin{eqnarray}
f(x_R) =
\frac{\alpha_W}{16\pi}\left[\frac{x_H}{x_R-x_H}\ln\left(\frac{x_R}{x_H}\right)
  + \frac{3x_Z}{x_R-x_Z}\ln\left(\frac{x_R}{x_Z}\right) \right] 
\end{eqnarray} 
with   $x_R   \equiv\hat{\mu}_R^2/m_W^2$,   $x_H\equiv   m_H^2/m_W^2$,
$x_Z\equiv m_Z^2/m_W^2$,  assuming $\mu_R =  \hat{\mu}_R{\bf 1}_3$ for
simplicity.  The functions $f(x_R)$  and $x_R  f(x_R)$ are  plotted in
Fig.~\ref{fig:2a}.

Thus, the  full neutrino mass  matrix in the  basis $\{\nu_{e,\mu,\tau
  L},S_{1,2,3 L}\}$ is given by
\begin{eqnarray}  
  \label{eq:Meffl}
{\cal M}_{\rm eff} = {\cal M}_{\rm eff}^{\rm tree}+{\cal M}_{\rm
  eff}^{\rm 1-loop} = \left(\begin{array}{cc}  
M_D\, \mu_R^{-1}\Big({\bf 1}_3+x_Rf(x_R)\Big)\, M_D^{\sf T} & 
M_D \mu_R^{-1}M_N^{\sf T} \\[2mm] 
M_N\mu_R^{-1}M_D^{\sf T} & M_N\mu_R^{-1}M_N^{\sf T} 
\end{array}\right)\; .
\end{eqnarray}
Since the loop-function $x_Rf(x_R)$ is  much smaller than unity for a wide
range of $\hat{\mu}_R$  as shown in Fig.~\ref{fig:2a}, we  can use the
unitary  matrix ${\cal  V}$ given  by~(\ref{eq:V0})  to perturbatively
block-diagonalize  the   full  mass  matrix   in~(\ref{eq:Meffl}),  as
follows:
\begin{eqnarray}
{\cal V}^{\sf T}{\cal M}_{\rm eff}{\cal V} &=& \left(\begin{array}{ccc}
\left({\bf 1}_3+\zeta\zeta^\dag\right)^{-1/2}\Delta M\left({\bf
  1}+\zeta^*\zeta^{\sf T}\right)^{-1/2} & & 
\left({\bf 1}_3+\zeta\zeta^\dag\right)^{-1/2}\Delta M\zeta^*\left({\bf
  1}+\zeta^{\sf T}\zeta^*\right)^{-1/2} \\ 
\left({\bf 1}_3+\zeta^\dag\zeta\right)^{-1/2}\zeta^\dag\Delta
M\left({\bf 1}+\zeta^*\zeta^{\sf T}\right)^{-1/2} & & 
M\mu_R^{-1}M^{\sf T}+ \left({\bf
  1}_3+\zeta^\dag\zeta\right)^{-1/2}\zeta^\dag\Delta
M\zeta^*\left({\bf 1}+\zeta^{\sf T}\zeta^*\right)^{-1/2}  
\end{array}\right)\nonumber\\ 
&\equiv & \left(\begin{array}{cc}
m_{11} & m_{12} \\
m_{12}^{\sf T} & M\mu_R^{-1}M^{\sf T}+m_{22}
\end{array}\right)\; .
\label{eq:block-Meff}
\end{eqnarray}  
Since $\|m_{12}\|,\|m_{22}\|\ll \|M\mu_R^{-1}M^{\sf T}\|$, we can make
the  usual   seesaw-like  approximation  to  obtain   the  block  mass
eigenvalues:
\begin{eqnarray}
  \label{eq:m1}
M_1 &\simeq& m_{11}-m_{12}\left(M\mu_R^{-1}M^{\sf T}\right)^{-1}
m_{12}^{\sf T}\; ,\\ 
   \label{eq:m2}
M_2 &\simeq& M\mu_R^{-1}M^{\sf T}\; .
\end{eqnarray}
It  is clear  from~(\ref{eq:m1}) that  the light  neutrino  mass block
eigenvalue is  proportional to the loop-correction  factor $\Delta M$,
and  vanishes as  $\Delta M\to  {\bf 0}$.  Note that  the  second term
in~(\ref{eq:m1})  represents an  effective two-loop  effect  of ${\cal
  O}((\Delta  M)^2)$, and next  to the  first term,  it can  be safely
ignored.

In order  to obtain  the active and  sterile components for  each mass
eigenvalue,  the  mass  matrix  in~(\ref{eq:block-Meff}) needs  to  be
further diagonalized into a  fully mass-diagonal form. Before doing so
for  the general  three-flavour  scenario in Section~\ref{general}, we  will  first consider  a
simple  one-flavour scenario for illustration purposes,  which would  help us  to  gain valuable
insight.

\section{The Single Flavour Case}\label{simple}

\begin{figure}
\centering
\includegraphics[width=7cm]{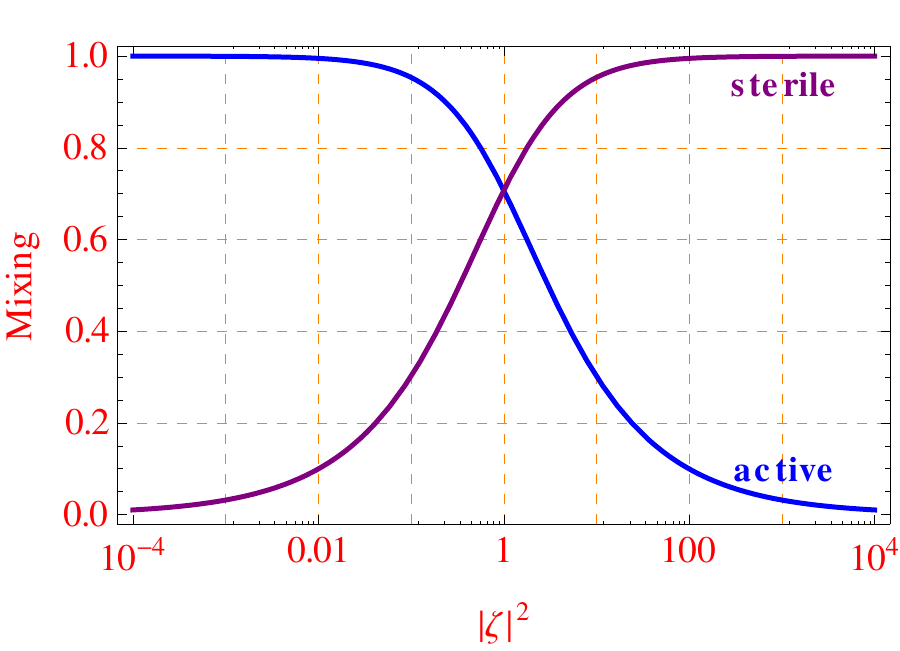}
\caption{The active and sterile components in the lighter neutrino
  state ${n}_1$ defined in~(\ref{eq:eigvec}) for the
  single flavour case.} 
\label{fig:1}
\end{figure}

In   this  case,   we  replace   all  the   matrices  in~(\ref{eq:m1})
and~(\ref{eq:m2}) by complex numbers.   Then, the mass eigenvalues are
simply given by
\begin{eqnarray}
m_1 =
\frac{x_Rf(x_R)m_N^2}{\hat{\mu}_R}\left(\frac{\zeta^2}{1+|\zeta|^2}\right)\;,
\qquad 
m_2 = \frac{m_N^2}{\hat{\mu}_R}(1+|\zeta|^2)\; ,
\end{eqnarray} 
while the mass eigenvectors are given by 
\begin{eqnarray}
  \label{eq:eigvec}
\left(\begin{array}{c}
n_1 \\ {n}_2 \end{array}\right) = {\cal V}^\dag\left(\begin{array}{c}
{\nu}_L \\ {\nu}_S \end{array}\right) 
= ({1}+|\zeta|^2)^{-1/2} \left(\begin{array}{c}
{\nu}_L-\zeta^* {S}_L \\
\zeta {\nu}_L + {S}_L
\end{array}\right)\; . 
\end{eqnarray}
From~(\ref{eq:eigvec}), we  note that  in the limit  $\|\zeta\|\ll 1$,
the  massless state  is  mostly the  active neutrino  ${\nu}_L$,
whereas in the limit $\|\zeta\|\gg  1$, the massless state is mostly
the   sterile  neutrino   ${S}_L$.   This   generic   property  is
illustrated in Fig.~\ref{fig:1}.

\begin{figure}
\centering
\includegraphics[width=7cm]{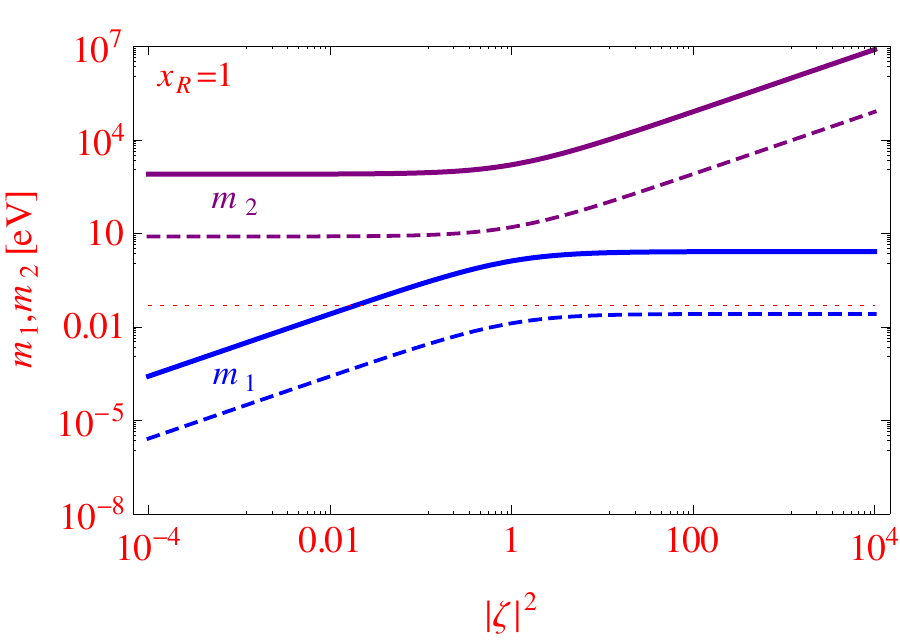}
\hspace{1cm}
\includegraphics[width=7cm]{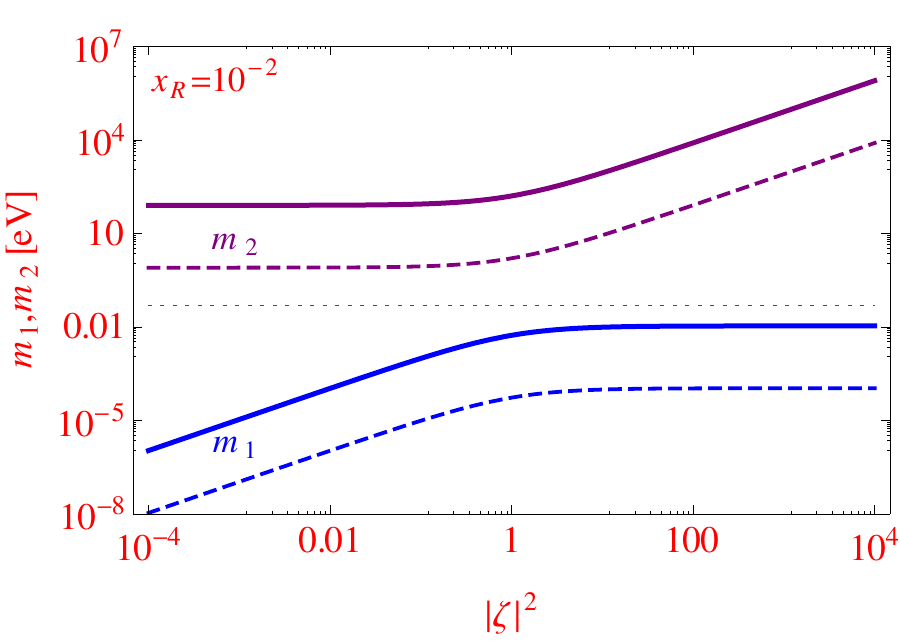}
\caption{The light neutrino mass eigenvalues $m_1$  and $m_2$ for the single flavour case with two different choices of $x_R=\mu_R^2/m_W^2$ and with $|m_N|=10^{-4}\mu_R$ (solid), $10^{-5}\mu_R$ (dashed). The
  horizontal red (dotted) line represents the tentative upper limit on the active neutrino mass scale.} 
\label{fig:3}
\end{figure}

The magnitudes of these  eigenvalues are shown in Fig.~\ref{fig:3}, as a function of $|\zeta|^2$  for two typical values of  the Majorana mass scale $\mu_R =  x_R m_W^2$. In each  case, we show the  values for two
choices of $|m_N|/\mu_R=10^{-4},10^{-5}$, given by the solid and
dashed  lines,  respectively. The  dotted  horizontal  line shows  the
experimental  upper limit  on  the active  neutrino  mass, assuming  a
normal hierarchy with vanishing $\nu_e$ mass. From Fig.~\ref{fig:3} we
see  that even  in this  simple  scenario, it  is possible  to have  a
sterile neutrino in the keV-mass range  to be a DM candidate (for
$|m_N|/\mu_R\gsim 10^{-4}$)  or in  the eV-range to  explain the
LSND  anomaly  (for  $|m_N|/\mu_R\lsim  10^{-5}$). In  both  the
cases,  we  must  have  $|\zeta|^2\ll  1$,  so  that  the  lighter  mass
eigenstate is  mostly the active neutrino  (cf. Fig.~\ref{fig:1}) with
mass in the sub-eV range.

\section{The Three Flavour Case}\label{general}

We  now turn  our  attention to  the  general case  with three  active
neutrinos $(\nu_e,\nu_\mu,\nu_\tau)$ and three light sterile neutrinos
$(\nu_{s_1},\nu_{s_2},\nu_{s_3})$.  In  this case, we  diagonalize the
full $6\times  6$ light neutrino mass matrix  in~(\ref{eq:Meffl}) by a
general unitary matrix~${\cal U}_6$:
\begin{eqnarray}
{\cal M}_{\rm eff}  = {\cal U}_6^{*} \left(\begin{array}{cc}
\widehat{M}_{123} & {\bf 0}_3\\
{\bf 0}_3 & \widehat{M}_{456}
\end{array}\right){\cal U}_6^\dag,
\label{eq:M-hat}
\end{eqnarray}
where    $\widehat{M}_{123}    =    {\rm    diag}(m_1,m_2,m_3)$    and
$\widehat{M}_{456}={\rm   diag}(m_4,m_5,m_6)$,  with  $m_{1,\ldots,6}$
being the mass eigenvalues of ${\cal M}_{\rm eff}$. Moreover,
\begin{eqnarray}
{\cal U}_6 = 
\left(\begin{array}{cccccc}
U_{e1} & U_{e2} & U_{e3} & U_{e4} & U_{e5} & U_{e6} \\
U_{\mu 1} & U_{\mu 2} & U_{\mu 3} & U_{\mu 4} & U_{\mu 5} & U_{\mu 6} \\
U_{\tau 1} & U_{\tau 2} & U_{\tau 3} & U_{\tau 4} & U_{\tau 5} & U_{\tau 6} \\
U_{s_11} & U_{s_12} & U_{s_13} & U_{s_1 4} & U_{s_15} & U_{s_16} \\
U_{s_21} & U_{s_22} & U_{s_23} & U_{s_24} & U_{s_25} & U_{s_26} \\
U_{s_31} & U_{s_32} & U_{s_33} & U_{s_34} & U_{s_35} & U_{s_36}
\end{array}\right)
\label{eq:U6u}
\end{eqnarray}
is  a $6\times  6$  unitary  matrix that  relates  the flavour  states
$\{\nu_e,\nu_\mu,\nu_\tau,\nu_{s_1},\nu_{s_2},\nu_{s_3}\}$ to the mass
eigenstates~$\{\nu_1,\nu_2, \ldots, \nu_6\}$.

As shown in  Appendix A, ${\cal U}_6$ can be  parametrized in terms of
15 Euler  angles and 10  Dirac phases\footnote{For simplicity,  we set
  all Majorana  phases to  zero.}.  This, together  with the  six mass
eigenvalues,   introduces   a   total   of  31   parameters   on   the
right-hand-side of~(\ref{eq:M-hat})  some of which  can be constrained
experimentally, as follows.  The only experimentally measured neutrino
parameters     so    far     are    the     three     mixing    angles
$\theta_{12},\theta_{23},\theta_{13}$   between   the  mostly   active
neutrinos, and the two  mass-squared differences $\Delta m^2_{21}$ and
$\Delta m^2_{31}$~\cite{pdg}. Also, there  is some hint for a nonzero
Dirac  $CP$  phase $\varphi_{13}$  from  the  global  analysis of  the
3-neutrino data~\cite{global,othergf}. For  our numerical analysis, we
use the latest best-fit values of these parameters as given in~\cite{global},
taking into account the reactor SBL data and assuming
a normal hierarchy:
\begin{eqnarray}
&&\theta_{12}=33.3^\circ\,,\quad
  \theta_{23}=40.0^\circ\,, \quad 
   \theta_{13}=8.6^\circ\,, \quad
  \varphi_{13}=1.67\pi,\nonumber\\ 
&& \Delta m^2_{21}= 7.50\times 10^{-5}~{\rm eV}^2\,,\quad
\Delta m^2_{31} =   2.47\times 10^{-3}~{\rm eV}^2\, . 
\end{eqnarray} 

On the  other hand, for the  mass and mixing  parameters involving the
sterile sector, there has been a number of recent analyses~\cite{kopp,
  giunti, donini,conrad,  fornengo}  involving one,  two  and
three sterile  neutrinos, denoted as (3+1),  (3+2) and (3+3)-scenarios
respectively.   Although  their  global best-fit  values for  any  particular
scenario  differ somewhat  from  each  other, they  all  agree on  the
conclusion  that there is a strong tension in the global data for the 
(3+1)-scenario~\cite{white, 3+1s}. This is mainly because 
the neutrino versus anti-neutrino and appearance versus disappearance data sets 
prefer two different values of $\Delta m^2_{41}$ for the 
(3+1)-scenario~\cite{conrad}. The 
neutrino versus anti-neutrino discrepancy is somewhat reduced in the (3+2)- and 
(3+3)-scenarios due to the additional $CP$-violating phases, thus leading to better global fits. However, the appearance versus disappearance tension still 
remains, and in particular, the 
fit to the latest MiniBooNE low-energy data is not improved by adding more 
sterile neutrinos. Thus due to the lack of a clear preference at the moment, 
we will consider benchmark cases for our model using both (3+1)- and 
(3+2)-fits, and will also comment on the (3+3)-fit.

There  also exist  upper bounds  on the  absolute neutrino  mass scale
which could in principle be used  to constrain the masses of the light
sterile  neutrinos  and  their  mixing  with the  active  ones.  These
constraints are briefly summarized below:
\begin{enumerate}

\item   The  effective  neutrino   mass  parameter   in  $\beta$-decay
  experiments:
\begin{eqnarray}
m_\beta = \sqrt{\sum_i |U_{ei}|^2 m_i^2}\;,
\end{eqnarray}
which determines the distortion of the electron energy spectrum due to
nonzero neutrino  mass and mixing.  The current  most stringent upper
limit on this  parameter is $m_\beta\leq 2.2$ eV~\cite{troitsk,mainz},
which  is still  compatible with  the  entire favoured  region of  the
global       fits~\cite{kraus}.        The       upcoming       KATRIN
experiment~\cite{katrin} with  estimated sensitivity reach,  as low as
0.2~eV, should be able to  probe a substantial fraction of the allowed
parameter space~\cite{esmaili}.

\item  The effective  neutrino mass  parameter in  neutrinoless double
  beta decay ($0\nu\beta\beta$) experiments:
\begin{eqnarray}
\langle m\rangle_{\beta\beta} = \left|\sum_i U_{ei}^2 m_i\right| =
\left|\sum_i |U_{ei}|^2 e^{i\alpha_i} m_i\right|\; ,
\label{0nubb}
\end{eqnarray}
where $\alpha_i$ are the  Majorana phases associated with the elements
$U_{ei}$.    The    strongest   bound    so   far   on    $\langle   m
\rangle_{\beta\beta}<0.26$~eV~\cite{0nubbex1,0nubbex2,0nubbex3, exo}     and
significant  improvements  are expected  to  take  place  in the  near
future~\cite{0nubbfuture}.    However,  since   the   Majorana  phases
in~(\ref{0nubb})   are  still   unknown,  it   is  possible   to  have
cancellations among  them~\cite{li-liu}, and hence,  the constraint on
$\langle m\rangle_{\beta\beta}$ may not apply for the sterile neutrino
sector.    For   our   benchmark   model   parameters   discussed   in
Section~\ref{benchmark},  the $0\nu\beta\beta$  contribution  from the
sterile sector is known to be negligible~\cite{pascoli}.

\item The  sum of the  neutrino masses, $\sum_i m_i$,  contributing to
  the total  energy density  in our Universe.   This parameter  can be
  constrained  from  the  analysis  of data  on  several  cosmological
  observables~\cite{pastor}, and  the current upper limits  are in the
  eV  range which  are in  tension with  the  sterile neutrino-favored
  region of the SBL  data. The cosmologically-derived bounds, however,
  depend on  the assumptions made on  the cosmic history  of the early
  Universe,  as well as  on the  corresponding cosmological  data set
  used,    besides    the    uncertainties   associated    with    the
  observables~\cite{fornengo,hamann,nucosmo}. It  might be possible to
  resolve some of these issues by analyzing the upcoming galaxy survey
  data~\cite{nucosmofuture}. It might also  be plausible to consider a
  deviation  from  the  standard  cosmological picture,  in  order  to
  accommodate the  eV-scale sterile neutrinos  as favoured by  the SBL
  data~\cite{white}. A  proper analysis of this  possibility is beyond
  the scope of this paper.  Hence, we do not consider the cosmological
  bounds for the rest of our discussion.
\end{enumerate}

In addition, the constraints on the active-sterile mixing from the 
non-unitarity  of  the  PMNS    mixing matrix~\cite{antusch}  need to  be 
taken  into account  as  well.  In the inverse seesaw model with 
$\mu_R={\bf 0}$ and $\mu_S\neq {\bf 0}$, the  non-unitarity effects are proportional to $\|\zeta\|^2$~\cite{nuty}. For the MRISM with $\mu_R\neq {\bf 0}$ and $\mu_S={\bf 0}$, the non-unitarity effects due to the mixing between the active and heavy sterile neutrinos in (\ref{eq:inverse1}) is of order 
$\|M_D\mu_R^{-1}\|^2$ 
which, as we will see in the following section, is very small ($\sim 10^{-9}$). 
This justifies our diagonalization procedure in (\ref{eq:M-hat}) using a unitary matrix ${\cal U}_6$. With  the  parametrization of
${\cal U}_6$  as given  in Appendix A,  the non-unitarity due to the mixing between the active and light sterile neutrinos is determined by $RR^\dag$  [cf.~(\ref{eq:nuty})]. Using a
combination of neutrino oscillation  data and unitarity constraints in
weak decays, the following allowed ranges for  $|V|$ have been 
derived~\cite{antusch}:
\begin{eqnarray}
|V| = \left(\begin{array}{ccc}
0.75-0.89 & 0.45-0.65 & < 0.20\\
0.19-0.55 & 0.42-0.74 & 0.57-0.82\\
0.13-0.56 & 0.36-0.75 & 0.54-0.82
\end{array}\right)\; .
\label{eq:Vuni}
\end{eqnarray} 
However, the constraints derived from weak decays are not directly 
applicable for the sterile neutrino masses below the electroweak 
scale as in our case. Also the constraints from Lepton Flavour Universality
which may  lead to measurable enhancements in  various flavour physics
observables~\cite{abada} are not applicable to our case with all the 
light sterile masses much below the MeV scale. We find that the numerical values of the 
active neutrino mixing matrix elements in both of 
our benchmark scenarios are well within the $1\sigma$ interval 
of the current global fit values from 3-neutrino oscillation 
data~\cite{global}: 
\begin{eqnarray}
|V| = \left(\begin{array}{ccc}
0.795-0.846 & 0.513-0.585 & 0.126-0.178\\
0.205-0.543 & 0.416-0.730 & 0.579-0.808\\
0.215-0.548 & 0.409-0.725 & 0.567-0.800
\end{array}\right).
\label{eq:VP}
\end{eqnarray} 
Note here that the constraints due to (\ref{eq:VP}) are actually stronger than 
those obtained from (\ref{eq:Vuni}).

We  conclude  this section  by  deriving  a  relationship between  the
general  unitary  matrix  ${\cal  U}_6$ that  fully  diagonalizes  the
effective    neutrino    mass     matrix    ${\cal    M}_{\rm    eff}$
in~(\ref{eq:Meffl})  and  the  unitary  matrix ${\cal  V}$  introduced
in~(\ref{eq:V}) which only  block-diagonalizes ${\cal M}_{\rm eff}$ as
in~(\ref{eq:block-Meff}).  This   relationship  can  be   obtained  by
rewriting~(\ref{eq:U6a}) as follows:
\begin{eqnarray}
{\cal U}_6 = \left(\begin{array}{cc}
{\bf 1}_3 & {\bf 0}_3\\
{\bf 0}_3 & U_0
\end{array}\right)
\left(\begin{array}{cc}
A & R \\
S & B 
\end{array}\right)
\left(\begin{array}{cc}
V_0 & {\bf 0}_3\\
{\bf 0}_3 & {\bf 1}_3
\end{array}\right) = 
{\cal V}\left(\begin{array}{cc}
{\bf 1}_3 & {\bf 0}_3\\
{\bf 0}_3 & U_0
\end{array}\right)\left(\begin{array}{cc}
V_0 & {\bf 0}_3\\
{\bf 0}_3 & {\bf 1}_3
\end{array}\right),
\end{eqnarray}
where ${\cal V}$ is identified as 
\begin{eqnarray}
{\cal V} = \left(\begin{array}{cc}
{\bf 1}_3 & {\bf 0}_3\\
{\bf 0}_3 & U_0
\end{array}\right)
\left(\begin{array}{cc}
A & R \\
S & B 
\end{array}\right)\left(\begin{array}{cc}
{\bf 1}_3 & {\bf 0}_3\\
{\bf 0}_3 & U^\dag_0
\end{array}\right)\equiv \left(\begin{array}{cc}
{\bf 1}_3 & {\bf 0}_3\\
{\bf 0}_3 & U_0
\end{array}\right) \widetilde{{\cal V}} \left(\begin{array}{cc}
{\bf 1}_3 & {\bf 0}_3\\
{\bf 0}_3 & U^\dag_0
\end{array}\right),
\end{eqnarray}
and   $\widetilde{{\cal   V}}$   has   an  exact   representation   as
in~(\ref{eq:V}) with the replacement $\zeta\to \widetilde{\zeta}=\zeta
U_0^*$.

\section{Benchmark Scenarios for Light and Superlight Sterile
  Neutrinos}\label{benchmark} 

\begin{figure}[h!]
\begin{center}
\includegraphics[width=7cm]{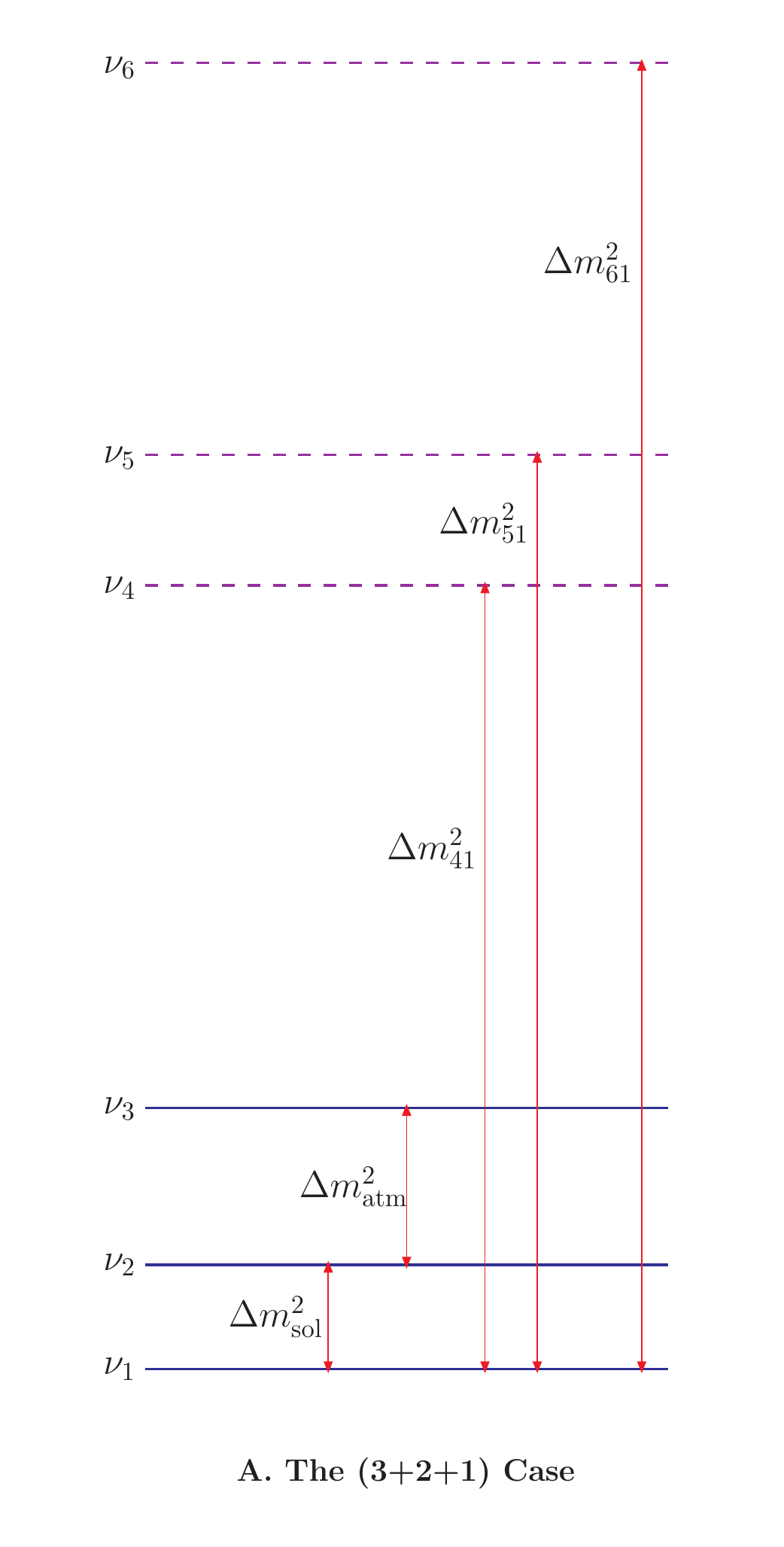}
\hspace{1cm}
\includegraphics[width=7cm]{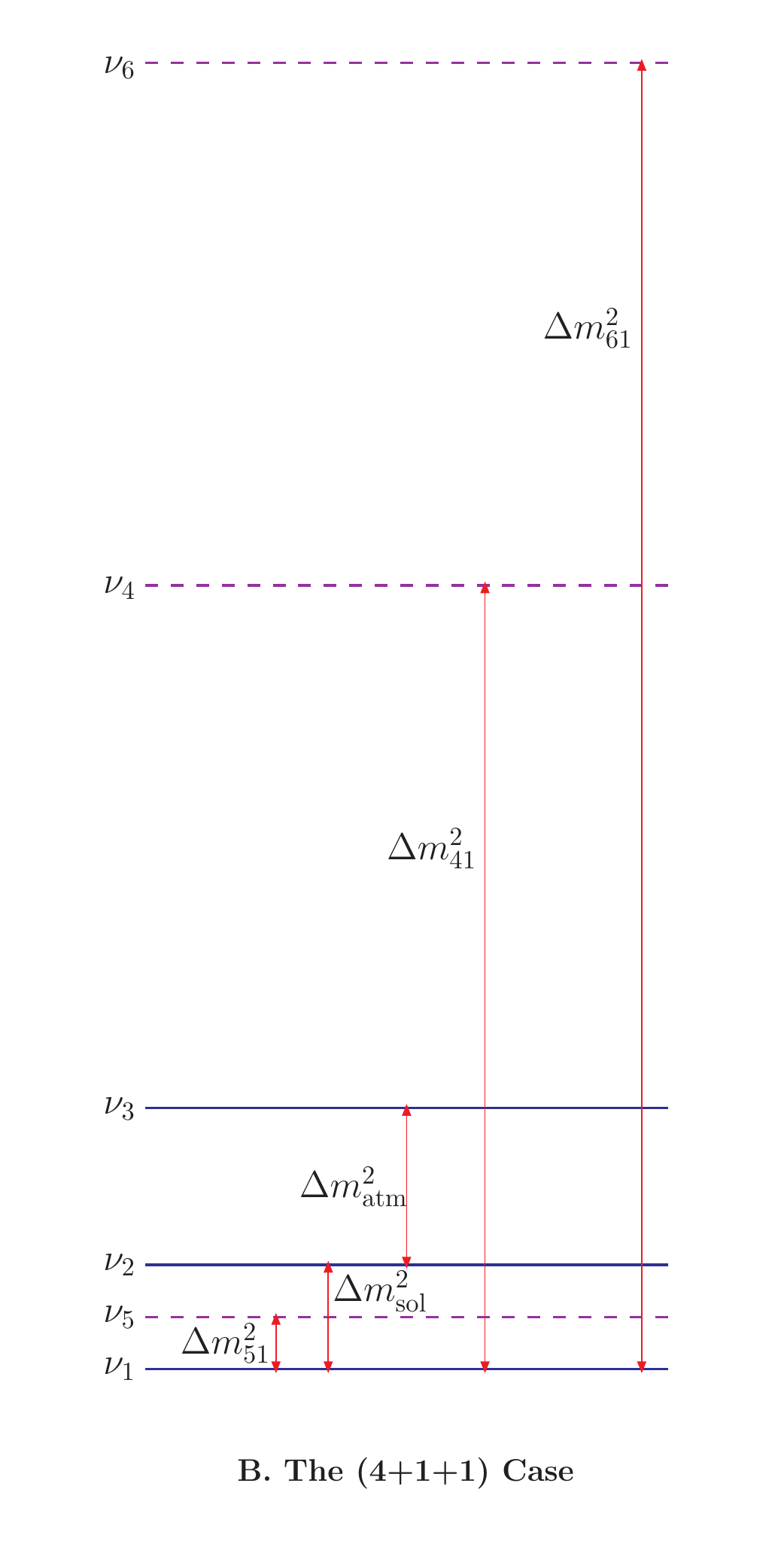}
\end{center}
\caption{The two benchmark scenarios discussed in Sections~\ref{321} and~\ref{422}. The solid (dashed) horizontal lines represent the mostly active (sterile) neutrino mass eigenstates. The masses are not to scale.}
\label{hierarchy}
\end{figure}

In  this section,  we  consider two  physically interesting  benchmark
scenarios as shown in Fig.~\ref{hierarchy} and comment  on  a  third one  considered  recently in  the
literature.  We  develop a method which  enables us to fit  some of the
model parameters for  each benchmark scenario, so as  to be compatible with 
the  SBL data. It also allows  us to  determine the
values for the remaining theoretical parameters that have not yet been
constrained by  the existing neutrino data  and could be probed  in future
experiments.

Employing the block form  of ${\cal U}_6$ in~(\ref{eq:U6a}), we obtain
from~(\ref{eq:M-hat})  the  following  consistency conditions  on  the
model parameters $M_D,M_N$ and $\mu_R$:
\begin{eqnarray}
  \widetilde{R}^*\widehat{M}_{123}\widetilde{R}^\dag +
U^*\widehat{M}_{456}U^\dag &=& M_N\mu_R^{-1}M_N^{\sf
  T},\label{eq:consty3}\\ 
V^*\widehat{M}_{123}\widetilde{R}^\dag + R^*\widehat{M}_{456}U^\dag
&=& M_D\mu_R^{-1}M_N^{\sf T},\label{eq:consty2}\\ 
V^*\widehat{M}_{123}V^\dag + R^*\widehat{M}_{456}R^\dag &=&
M_D\,\mu_R^{-1} \Big({\bf 1}_3+x_Rf(x_R)\Big)\, M_D^{\sf
  T}\; \label{eq:consty1}. 
\end{eqnarray}
Using the best-fit values for the mass and mixing parameters discussed in Section~\ref{general}, we  can derive constraints  on the MRISM parameter  space.  For
simplicity, we  assume normal hierarchy\footnote{The  case of inverted
  hierarchy can  be worked out in  a similar way. For  a discussion of
  all  possible  scenarios  with  two  and  three  sterile  neutrinos,
  see~\cite{goswami}.}  for the  active  neutrinos, i.e.,  $m_1<m_2\ll
m_3$, with $m_1=0$.
%%%%%%%%%%%%%%%%%%%%%%%%%%%%
\subsection{The (3+2+1) Case}\label{321}
In  this scenario,  we  assume  that all  three  active neutrinos  are
lighter than the three light  sterile neutrinos, as shown schematically in Fig.~\ref{hierarchy}A. Among the three light
sterile neutrinos, we  assume one to be of  order keV-scale to account
for a warm/cold DM candidate and/or pulsar kicks~\cite{kusenko}.  Following
the analysis  of~\cite{AdG06}, we take the largest  mass eigenvalue to
be  $m_6=1$~keV and the  mixing parameters  $|U_{\alpha 6}|  = 7\times
10^{-3}$,  for $\alpha  =  e,\,\mu,\,\tau$, which  implies the  mixing
angles $\theta_{i6}=0.4^\circ$,  for $i=1,2,3$ in~(\ref{eq:U6a}).  We also  choose $\varphi_{i6}=0$  for  $i=1,2,3,4$ since these phases do not affect the SBL data significantly.  The
other two light sterile neutrinos  are assumed to have eV-scale masses
in   order  to   explain  the   LSND+MiniBooNE  results   and  reactor
anomalies~\cite{kopp,giunti, donini, conrad, fornengo}. Thus we have a 
simple extension of the (3+2)-models discussed in the literature which can now 
account for the DM as well. 

\begin{figure}[t]
\centering
\includegraphics[width=7cm]{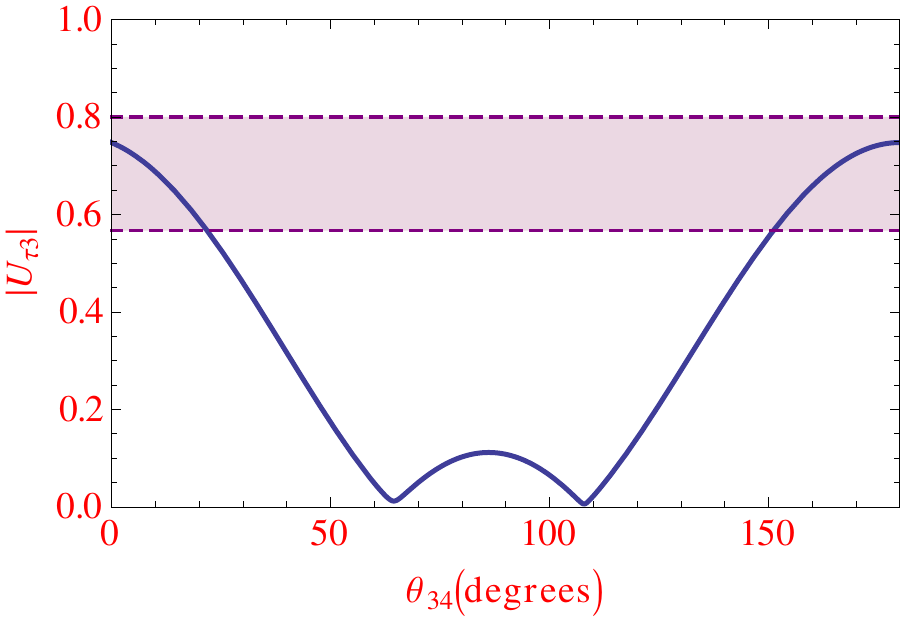}
\caption{The magnitude of the mixing element $U_{\tau 3}$ as a
  function of the mixing angle $\theta_{34}$. The shaded region shows the
  experimentally allowed range.} 
\label{fig:4}
\end{figure}

Since
the third sterile state is much heavier than the other two and its mixing effects are  very small,  we can still use the global analysis of the (3+2)-scenario.  From the latest global fit of~\cite{conrad} including the most recent MiniBooNE data\footnote{The allowed region in the $\Delta m^2_{41}-\Delta m^2_{51}$ plane from the global fit of~\cite{conrad} is somewhat different from the other recent global fits~\cite{kopp, giunti, donini, fornengo}. This is mainly because the disappearance data sets used in~\cite{conrad} prefer a medium $\Delta m^2_{41}$ (0.92 eV$^2$) and a high $\Delta m^2_{51}$ (18 eV$^2$) whereas the appearance data sets prefer a low $\Delta m^2_{41}$ (0.31 eV$^2$) and a medium $\Delta m^2_{51}$ (1.0 eV$^2$). This incompatibility is expected to be addressed soon after more MiniBooNE neutrino data become available.},  we get the best-fit values of $\Delta  m^2_{41}  =
0.92~{\rm eV}^2,~\Delta m^2_{51} = 17~{\rm eV}^2$, and
\begin{eqnarray}
\label{glst}
&& |U_{e4}| = 0.15\,,\quad 
|U_{\mu 4}| = 0.13\,,\quad 
|U_{e5}| = 0.069\,, \quad
|U_{\mu 5}|=0.16\,, \quad
\phi_{54} \equiv {\rm arg}(U_{e5}U^*_{\mu 5}U^*_{e4}U_{\mu 4}) = 1.8\pi\,,
\end{eqnarray} 
which imply  
\begin{eqnarray}
\label{eq:3+2bf}
&&\theta_{15} = 4.0^\circ\,,\quad 
\theta_{25} = 9.2^\circ\,, \quad 
\theta_{14} = 8.6^\circ\,,\quad 
\theta_{24} = 7.7^\circ\,,\quad
\phi_{54} \simeq
(\varphi_{14}-\varphi_{24})+(\varphi_{25}-\varphi_{15})=1.8\pi \, .
\end{eqnarray}
Note  that   the  only  experimentally  relevant   $CP$-phase  in  the
(3+2)-case   is  the   combination  $\phi_{54}$,   which   enters  the
$\bar{\nu}_\mu  \to  \bar{\nu}_e$  transition  probability.  Thus,  we
cannot determine all the 5 Dirac phases in the (3+2)-sector. We assume
for        simplicity        that        the       Dirac        phases
$\varphi_{24},\varphi_{15},\varphi_{25},\varphi_{35}$      are     all
zero\footnote{We find that allowing a  nonzero value for any of these
  phases   does   not  change   our   fit   significantly.}  so   that
$\phi_{54}\simeq  \varphi_{14}$.  We   also  take  the  mixing  angles
$\theta_{34}$  and  $\theta_{35}$  to   be  equal  which  can  now  be
constrained from the allowed range  of values for $|U_{\tau 3}|$ since
in the  PMNS parametrization this  element depends {\it only} on  the mixing
angles  $\theta_{23}$ and  $\theta_{13}$, whose  values are  now known
experimentally~\cite{pdg}. This is shown in Fig.~\ref{fig:4}, where we
plot      the      dependence      of      $|U_{\tau      3}|$      on
$\theta_{34}=\theta_{35}$.  The  allowed range  of  $|U_{\tau 3}|$  as
given   in~(\ref{eq:VP})  is   shown  as   the   shaded
region.   Thus,   we   see   that   in   the   first   quadrant   only
$\theta_{34}<22.1^\circ$ is allowed.

%%%

While $\theta_{34}$  and $\theta_{45}$ are constrained to  be small as
shown   in   Fig.~\ref{fig:4},   the   other  three   mixing   angles,
viz.  $\theta_{45},\theta_{46}$ and  $\theta_{56}$,  parametrizing the
mixing effects in the purely  sterile sector are mostly insensitive to
oscillation experiments. However, using the consistency conditions for
our model  parameters given by (\ref{eq:consty3})--(\ref{eq:consty1}),
we  can {\it  predict}  their  best-fit values  along  with the  model
parameters satisfying  these conditions.

\begin{figure}[h!]
\begin{center}
\includegraphics[width=8cm]{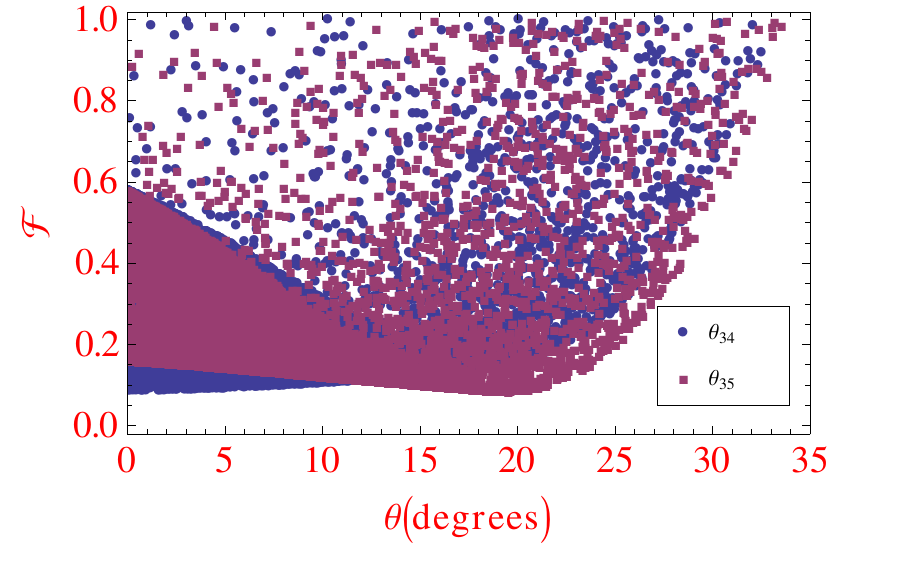}\hspace{0.0cm}
\includegraphics[width=8cm]{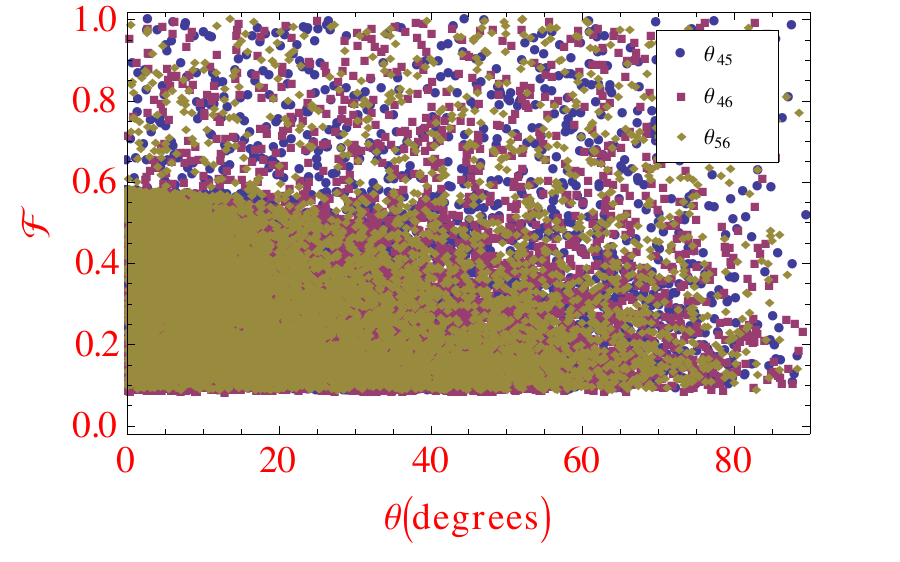}\\ \vspace{0.1cm}
\includegraphics[width=7cm]{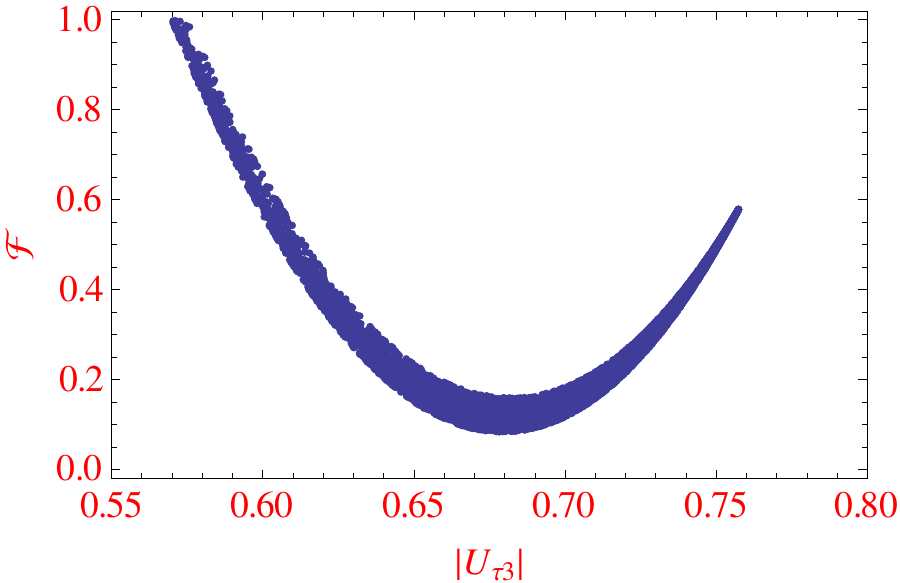}
\end{center}
\caption{The function ${\cal F}$ defined in~(\ref{eq:F}) versus the active-sterile neutrino mixing parameters for the (3+2+1) case.}
\label{fig:3p2p1}
\end{figure}

Our approach to data fitting may be described as follows.  For a given
set       of       values       for      the       mixing       angles
$\{\theta_{34},~\theta_{35},~\theta_{45},~\theta_{46},~\theta_{56}\}$
and  the scale  $\hat{\mu}_R$,  we  solve for  the  mass matrix  $M_N$
exactly using~(\ref{eq:consty3})  for a symmetric  structure of $M_N$,
i.e., for $M_N=M_N^{\sf  T}$. Then using~(\ref{eq:consty2}), we obtain
the input values  for the elements of the mass  matrix $M_D$, which we
use to  check the  validity of~(\ref{eq:consty1}). More  precisely, we
minimize the dimensionless function
\begin{eqnarray}
{\cal F} = \frac{\left\|\Big(V^*\widehat{M}_{123}V^\dag +
  R^*\widehat{M}_{456}R^\dag\Big)-\Big[ M_D\,\mu_R^{-1} \Big({\bf
      1}_3+x_Rf(x_R)\Big)\,M_D^{\sf
    T}\Big]\right\|}{\left\|V^*\widehat{M}_{123}V^\dag +
  R^*\widehat{M}_{456}R^\dag\right\|}\ ,  
\label{eq:F}
\end{eqnarray}
with       respect      to       the       set      of       variables
$\{\theta_{34},~\theta_{35},~\theta_{45},~\theta_{46},~\theta_{56}\}$
while  requiring that  the  elements $U_{\tau  i}$  for $i=1,2,3$  lie
within the allowed range given in~(\ref{eq:VP}). For illustration, our results for $x_R=1$ case are shown in Fig.~\ref{fig:3p2p1} and the best-fit values of the mixing angles that minimize the function ${\cal F}$ given by~(\ref{eq:F}) are 
\begin{eqnarray}
\theta_{34}= 0.1^\circ\,,\quad 
\theta_{35}= 19.6^\circ\,,\quad
\theta_{45}=26.4^\circ\,,\quad
\theta_{46}=12.9^\circ\,,\quad
\theta_{56}=54.8^\circ\,,
\end{eqnarray} 
which yields for the active neutrino mixing matrix 
\begin{eqnarray}
|V| = \left(\begin{array}{ccc}
0.8150 & 0.5354 & 0.1475\\
0.4793 & 0.5881 & 0.6182\\
0.3032 & 0.5826 & 0.6783
\end{array}\right),
\label{eq:VNa}
\end{eqnarray}
and the model parameter values
\begin{eqnarray}
M_D &=& \left(\begin{array}{ccc}
0.0605-0.0210i & 0.0529+0.0105i & 0.0127-0.0066i\\
0.0839+0.0002i & 0.0743 & -0.0270-0.0001i\\
0.0941+0.0003i & 0.1321-0.0002i & -0.1131+0.0001i
\end{array}\right)~{\rm MeV},\\
M_N &=& \left(\begin{array}{ccc}
0.7508+0.0004i & 1.5916+0.0001i & 0.9986-0.0003i\\
1.5916+0.0001i & 5.8432-0.0002i & 3.7990+0.0002i\\
0.9986-0.0003i & 3.7990+0.0002i & 3.1780-0.0002i
\end{array}\right)~{\rm MeV}\; .
\end{eqnarray}
with $\|\zeta\|=0.44$. Note that this leads to a very small non-unitary effect due to the mixing of the active neutrinos with the heavy sterile states: $\|M_D\mu_R^{-1}\|\sim 10^{-9}$. On the other hand, the non-unitarity due to the mixing with the light sterile states is well within the $1\sigma$ uncertainty in the active neutrino mixing matrix, as can be seen by comparing (\ref{eq:VP}) with (\ref{eq:VNa}).  
%%%%%%%%%%%%%%%%%%%%%%%%%%
\subsection{The (4+1+1) Case}\label{422}
In this case,  we assume that one of the sterile  neutrinos is in the keV-range to account for the DM as in the previous case. Another sterile state is 
assumed to be in the eV-range to explain the SBL data. Thus, this can be 
regarded as a simple extension of the (3+1)-models discussed in the literature. The third sterile state is superlight  (well below the  eV range), having  a weak mixing  to the
active  sector, so that  it does  not contribute  much to  the SBL 
data.  This scenario is shown schematically in Fig.~\ref{hierarchy}B. To the best of our knowledge, an explicit model with such a superlight sterile neutrino in the vicinity of the active neutrinos while being consistent with the SBL data has not been considered in the literature so far. Here, we  would like  to explore  such a  possibility  and its experimental prospects.

\begin{figure}[h!]
\centering
\includegraphics[width=7cm]{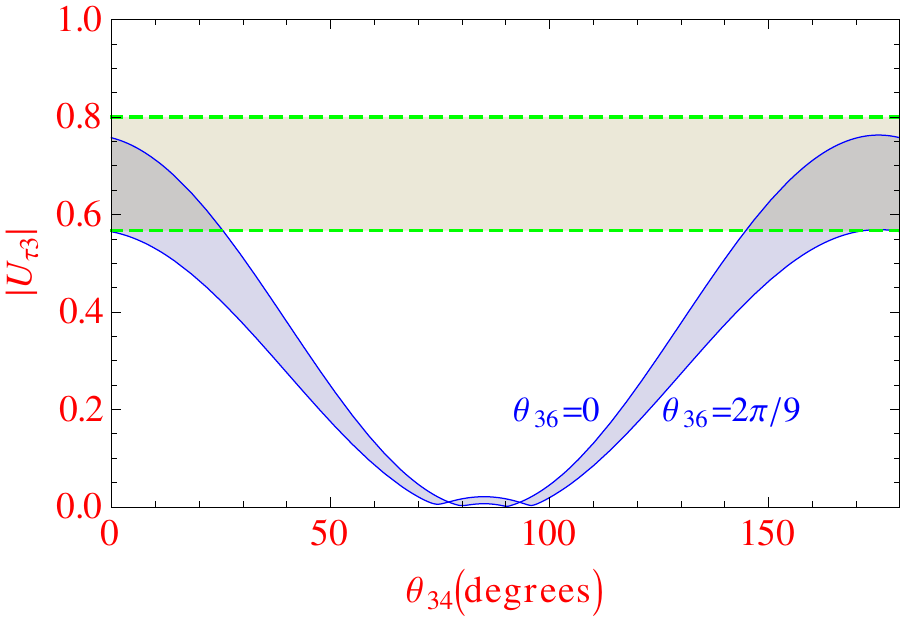}
\hspace{1cm}
\includegraphics[width=7cm]{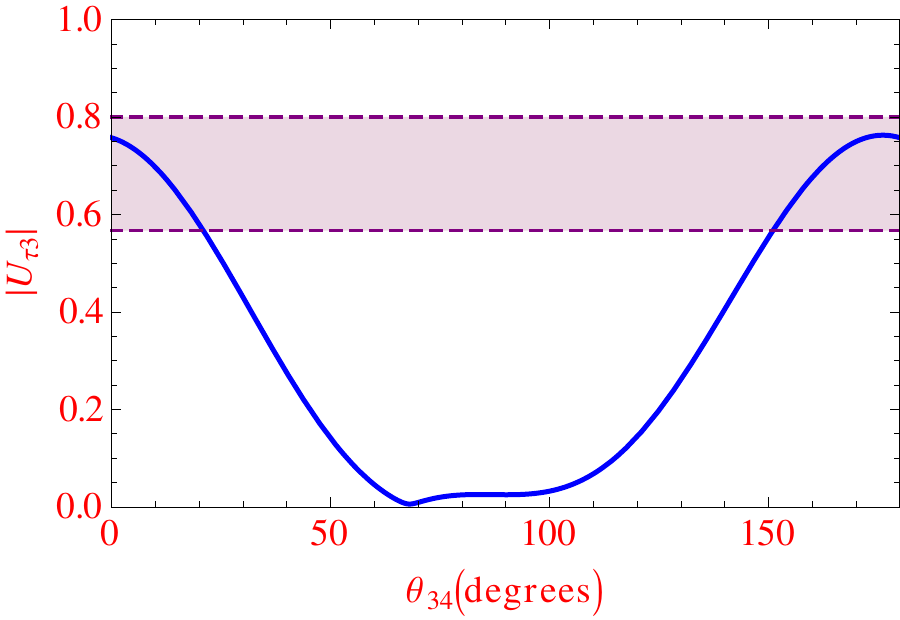}
\caption{The magnitude of the mixing element $U_{\tau 3}$ as a
  function of the mixing angle $\theta_{34}=\theta_{35}$ with
  $\theta_{36}$ as a free parameter (left panel) and with
  $\theta_{36}=\theta_{34}$ (right panel). The horizontal shaded
  region shows the experimentally allowed range.}
\label{fig:5}
\end{figure}

A superlight sterile state  in the vicinity of the electron  neutrino was considered in~\cite{smirnov} with $\Delta m^2_{61}\sim
(0.2-2)\times 10^{-5}~{\rm eV}^2$ and the mixing angle $\sin^2
2\theta_{16}\sim 10^{-3}$ in order to explain the absence of an upturn at low energies in the solar neutrino data. They also required the
mixing $|U_{s_3 3}|^2\sim 0.02-0.2$ in order to have a significant
production of this superlight sterile state in the early Universe,
which could generate the additional effective relativistic degree of
freedom, $\Delta N_{\rm eff} \sim 1$ as observed from the recent
Wilkinson Microwave Anisotropy Probe (WMAP) data~\cite{wmap}~\footnote{Note that in our case, the eV-scale sterile neutrino could also contribute to $\Delta N_{\rm eff}$, depending on its thermal history at the Big Bang Nucleosynthesis epoch.}, provided the mixing is large: $|U_{s_3 3}|^2\sim
0.1-0.2$. For illustration, let us consider a similar situation as
in~\cite{smirnov}, with
\begin{eqnarray}
\Delta m^2_{51}=2\times 10^{-5}~{\rm eV}^2\,,\qquad
\theta_{15}=\theta_{25}=2.0^\circ\;, 
\end{eqnarray}
while for the eV-scale sterile state, we use the best-fit value for the (3+1)-scenario~\cite{conrad}:
\begin{eqnarray}
\Delta m^2_{41}=0.92~{\rm eV}^2, \, |U_{e4}|=0.15,\, |U_{\mu 4}|=0.17.
\end{eqnarray}
Also, we  assume for  simplicity that  the Dirac
phases
$\varphi_{24},\varphi_{15},\varphi_{25},\varphi_{35},\varphi_{16},\varphi_{26},\varphi_{36},\varphi_{46}$
are all zero. The allowed range of values for $|U_{\tau 3}|$ will then
constrain  the values of  the mixing  angles $\theta_{34},\theta_{35}$
and $\theta_{36}$ to be  small. For $\theta_{34}=\theta_{35}$, this is
shown in the left-panel of  Fig.~\ref{fig:5}  where we have  shown the allowed
range  of $\theta_{36}=[0,2\pi/9]$. On  the other  hand, if  we choose
$\theta_{34}=\theta_{35}=\theta_{36}$,  then  the  mixing  angles  are
constrained to  be $<22.1^\circ$, as in
the previous scenario. This is shown in the right panel of Fig.~\ref{fig:5}.

\begin{figure}[h!]
\begin{center}
\includegraphics[width=8cm]{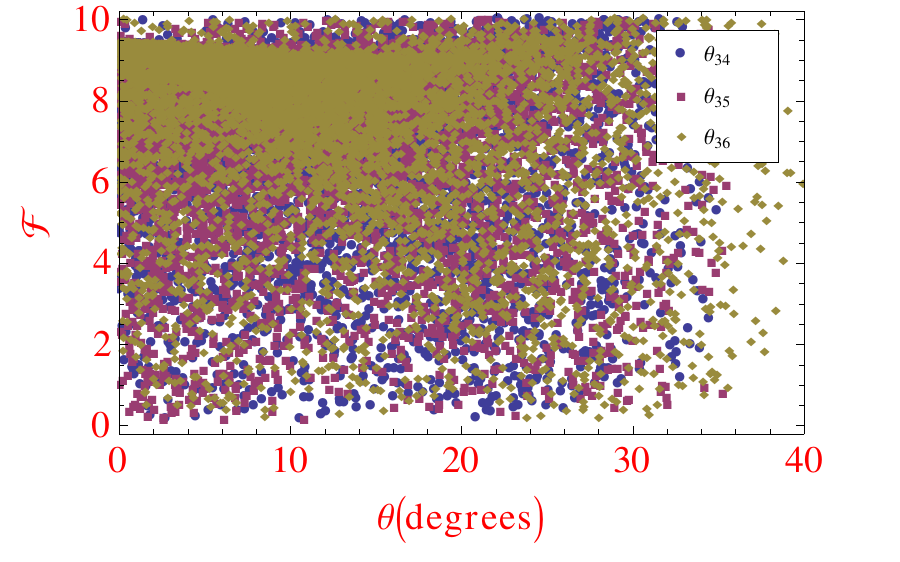}\hspace{0.0cm}
\includegraphics[width=8cm]{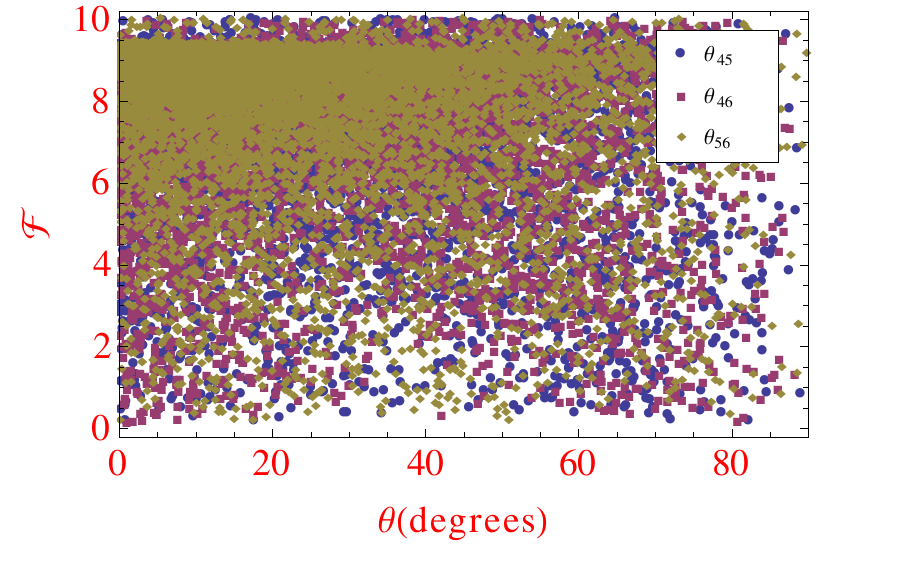}\\ \vspace{0.1cm}
\includegraphics[width=7cm]{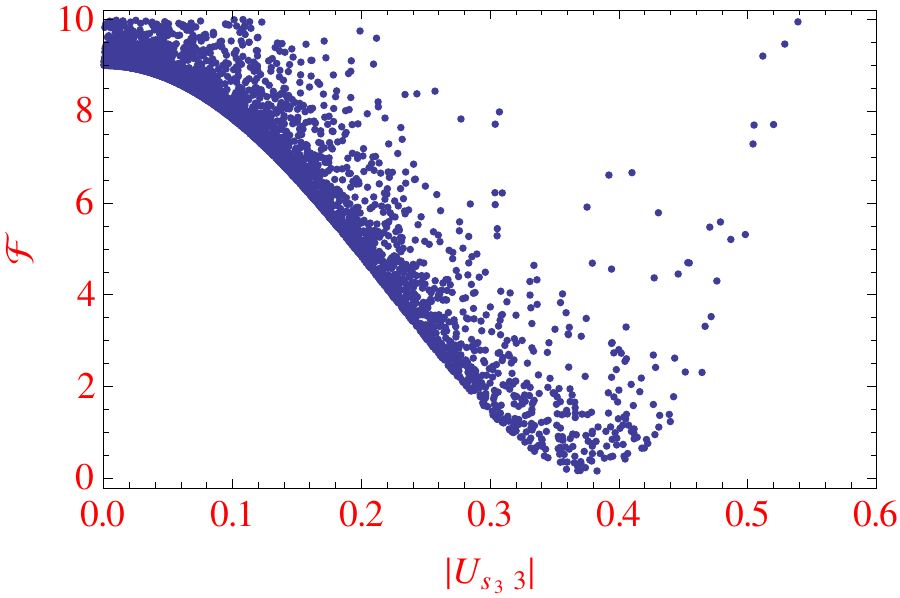}\hspace{0.5cm}
\includegraphics[width=7cm]{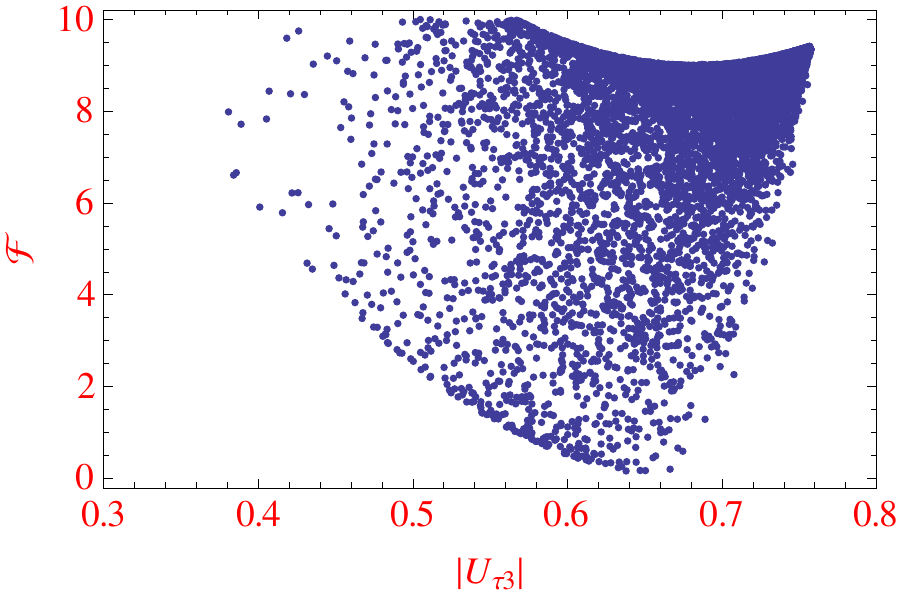}
\end{center}
\caption{The function ${\cal F}$ defined in~(\ref{eq:F}) versus the active-sterile mixing parameters for the (4+1+1) case.}
\label{fig:4p2}
\end{figure}

Using  the  consistency  conditions  for our  model  parameters  given
by~(\ref{eq:consty3})--(\ref{eq:consty1}) , we can derive the best-fit
values      for     the     set      of     six      mixing     angles
$\{\theta_{34},~\theta_{35},~\theta_{36},~\theta_{45},~\theta_{46},~\theta_{56}\}$,
as well as the corresponding  model parameter values. To achieve this,
we minimize the  function defined in~(\ref{eq:F}) while simultaneously
requiring that  the elements $|U_{\tau i}|$ for  $i=1,2,3$ fall within
the  allowed  range  of  $|V_{\tau 3}|$  given  in~(\ref{eq:VP}).   In
addition, we require that the element $|U_{s_3 3}|^2$ should be within
$0.1-0.2$, as  discussed above. For $x_R=1$, our results are shown in Fig.~\ref{fig:4p2} and the best-fit values for
the mixing angles which minimize the function ${\cal F}$ defined in~(\ref{eq:F}) are
\begin{eqnarray}
\theta_{34}= 10.6^\circ\,,\quad
\theta_{35}=10.8^\circ\,,\quad
\theta_{36}=23.8^\circ\,,\quad
\theta_{45}=82.3^\circ\,,\quad
\theta_{46}=1.0^\circ\,,\quad
\theta_{56}=50.8^\circ\,,
\end{eqnarray}
with $|U_{s_33}|^2=0.16$ and 
\begin{eqnarray}
|V| = \left(\begin{array}{ccc}
0.8165 & 0.5364 & 0.1477\\
0.4788 & 0.5948 & 0.6221\\
0.2730 & 0.5498 & 0.6380
\end{array}\right) \, .
\label{VNb}
\end{eqnarray}
The corresponding model parameters are
\begin{eqnarray}
M_D &=& \left(\begin{array}{ccc}
0.0075+0.0021i & 0.2210+0.0154i & 0.2239-0.0188i\\ 
-0.0072+0.0002i & 0.2129 &  0.2337 \\ 
0.0439-0.0002i & 2.76678 - 0.0001i & 2.3209+0.0002i  
\end{array}\right)~{\rm MeV},\\
M_N &=& \left(\begin{array}{ccc} 
0.0258 & 0.0910 & 0.1060\\
0.0910 & 5.0303 & 3.8820 \\
0.1060 & 3.8820 & 3.4258 
\end{array}\right)~{\rm MeV}. 
\end{eqnarray}
with $\|\zeta\|=1.45$. Again in this case, we find that the non-unitarity effects are small, as can be seen by comparing (\ref{eq:VP}) with (\ref{VNb}). 

\subsection{The (3+3) Case}

Finally, we  would like  to comment  on the case  where all  the three
sterile neutrinos are in  the eV-range  and heavier than  the active ones.  It was shown in~\cite{conrad} that this scenario has the highest compatibility between
the   neutrino   and   anti-neutrino   data   sets, but still has a poor compatibility between the appearance and disappearance data sets.  They obtained  the following  best-fit values  for the
experimentally relevant 12 parameters in the (3+3)-case:
\begin{eqnarray}
&&\Delta m^2_{41}=0.90~{\rm eV}^2\,,\quad
\Delta m^2_{51}=17~{\rm eV}^2\,,\quad
\Delta m^2_{61}=22~{\rm eV}^2\,,\nonumber\\
&&|U_{\mu 4}|=0.12\,,\quad
|U_{e4}|=0.11\,, \quad
|U_{\mu 5}|=0.17\,, \quad
|U_{e5}|=0.11\,,\quad
|U_{\mu 6}|=0.14\,,\quad
|U_{e6}|=0.11,\nonumber\\
&&\phi_{54}\equiv {\rm arg}(U_{e5}U^*_{\mu 5}U^*_{e4}U_{\mu
  4})=1.6\pi\,,\quad 
\phi_{64}\equiv {\rm arg}(U_{e6}U^*_{\mu 6}U^*_{e4}U_{\mu
  4})=0.28\pi\,,\quad
\phi_{65}\equiv {\rm arg}(U_{e6}U^*_{\mu 6}U^*_{e5}U_{\mu
  5})=1.4\pi\; .\qquad\label{eq:3+3gvp}
\end{eqnarray}
For the parametrization of the unitary matrix discussed in Appendix A,
this yields
\begin{eqnarray}
&&\theta_{16} = 6.3^\circ\,,\quad
\theta_{26}=8.1^\circ\,,\quad
\theta_{15}=6.3^\circ\,,\quad
\theta_{25}=9.9^\circ\,,\quad
\theta_{14}=6.6^\circ\,,\quad
\theta_{24}=7.1^\circ\,,\label{small}\\
&&\phi_{54}\simeq
(\varphi_{14}-\varphi_{24})+(\varphi_{25}-\varphi_{15})\,, \quad 
\phi_{64}\simeq
(\varphi_{14}-\varphi_{24})+(\varphi_{26}-\varphi_{16})\,,
\quad 
\phi_{65}\simeq
  (\varphi_{15}-\varphi_{25})+(\varphi_{26}-\varphi_{16})\,
 .\qquad
\label{eq:3+3g}
\end{eqnarray}
Thus, we find that the three experimentally relevant Dirac $CP$ phases
for the (3+3) scenario are {\it not}  independent of each other, and only two of them are independent, as already noted in~\cite{maltoni}. Hence, we  cannot  obtain a  solution for  the
phases     in~(\ref{eq:3+3g})    satisfying    the     values    given in~(\ref{eq:3+3gvp}),  which  were presumably obtained  by  assuming  that they  are independent parameters. Note that it is also very difficult to make 
the (3+3)-scenario with the best-fit $\Delta m^2$ values given by 
(\ref{eq:3+3gvp}) compatible with a standard cosmological model.   

\section{Conclusions}\label{conclude}

We  have discussed  the  possibility of  having  light and  superlight
sterile neutrinos  in the recently proposed  Minimal Radiative Inverse
Seesaw  Model.   In the  limit  $\|\mu_R\|\gg \|M_D\|,\|M_N\|$,  after
integrating  out  the  heavy  singlet  states  with  masses  of  order
$\|\mu_R\|$, we are left with three light sterile neutrinos along with
the  three usual active  neutrinos. We  have considered  two benchmark
scenarios for the mass hierarchy of these light sterile states. In the
first  scenario, denoted  as the  (3+2+1)-scenario, one  of  the light
sterile states  has mass in the  keV-range and very  small mixing with
the active neutrinos  to account for the Dark  Matter in the Universe,
whilst the other two sterile  states are in the eV-range with non-zero
mixing   with  the  active   states,  as   required  to   explain  the
LSND+MiniBooNE+reactor neutrino data. 

In  this   article,  we   have  also  discussed   another  potentially
interesting scenario, denoted as the (4+1+1)-scenario, 
with one superlight sterile neutrino state almost
degenerate in mass with the solar neutrinos, within the context of the
Minimal  Radiative  Inverse  Seesaw  Model.  This  superlight  sterile
neutrino may mix weakly with  both solar and atmospheric neutrinos and
might  give rise  to  observable  effects in  the  current and  future
neutrino data. Such  superlight sterile neutrinos may also provide
an explanation for the extra radiation observed in the Universe. Moreover, this scenario is more compatible with the cosmological constraints on the sum of the neutrino masses. 

Finally, we note that both of these scenarios discussed here constitute 
{\it   complete} extensions  of the Standard  Model for neutrino  masses and
Dark Matter,  while being  consistent with all  the existing  data. In
addition, the heavy singlet neutrinos  in this model with a degenerate
mass spectrum could be  used to explain the observed matter-antimatter
asymmetry  in   the  Universe,  through  the   mechanism  of  resonant
leptogenesis.

\begin{acknowledgments}
We acknowledge  helpful discussions with Justin  Evans, Manimala Mitra
and Alexei Smirnov.  We also  thank Christina Ignarra and He Zhang for
useful correspondence and suggestive comments.  This work is supported
by  the   Lancaster-Manchester-Sheffield  Consortium  for  Fundamental
Physics  under STFC  grant  ST/J000418/1. In  addition, AP  gratefully
acknowledges  partial  support by  a  IPPP  associateship from  Durham
University.
\end{acknowledgments}

\appendix

\section{Parametrization of a general unitary matrix}\label{app}

Following~\cite{harari}, we write a general $n\times n$ unitary matrix as
\begin{eqnarray}
{\cal U}_n = (\omega_{n-1,n}\omega_{n-2,n}\ldots
\omega_{1,n})(\omega_{n-2,n-1}\ldots \omega_{1,n-1})\ldots
(\omega_{2,3}\omega_{1,3})\omega_{1,2}\; ,
\label{eq:Un}
\end{eqnarray}     
where, for $j-i=1$, the rotation matrices $\omega_{i,j}$ are real:
\begin{eqnarray}
\omega_{i,j} = \left(\begin{array}{cccccccccc}
1 & & & & & & & & & \\
  & \cdot & & & & & & & & \\
  & & \cdot & & & & & & & \\
  & & & 1 & & & & & & \\
  & & & & \cos\theta_{ij} & \sin\theta_{ij} & & & &\\
  & & & & -\sin\theta_{ij} & \cos\theta_{ij} & & & & \\
  & & & & & & 1 & & &  \\
  & & & & & & & \cdot & &  \\
  & & & & & & & & \cdot &  \\
  & & & & & & & & & 1
\end{array}\right)~~({\rm for}~(j-i)=1),
\label{eq:omega1}
\end{eqnarray}
and for $j-i\geq 2$, the rotation matrices are complex:
\begin{eqnarray}
\omega_{i,j} = \left(\begin{array}{cccccccccccccc}
1 & & & & & & & & & & & & & \\
  & \cdot & & & & & & & & & & & & \\
  & & \cdot & & & & & & & & & & & \\
  & & & 1 & & & & & & & & & & \\
  & & & & \cos\theta_{ij} & 0 & \cdot & \cdot & 0 &
\sin\theta_{ij}e^{-i\varphi_{ij}} & & & &\\ 
  & & & & 0 & 1 & & & & 0 & & & & \\
  & & & & \cdot & & \cdot & & & \cdot & & & &\\
  & & & & \cdot & & & \cdot & & \cdot & & & &\\
  & & & & 0 & & & & 1 & 0 & & & & \\
  & & & & -\sin\theta_{ij}e^{i\varphi_{ij}} & 0 & \cdot & \cdot & 0 &
\cos\theta_{ij} & & & & \\ 
  & & & & & & & & & & 1 & & &  \\
  & & & & & & & & & & & \cdot & &  \\
  & & & & & & & & & & & & \cdot &  \\
  & & & & & & & & & & & & & 1
\end{array}\right)~~({\rm for}~(j-i)\geq 2),
\label{eq:omega2}
\end{eqnarray}

For our case of our interest here, i.e., for $n=6$,~(\ref{eq:Un}) becomes
\begin{eqnarray}
{\cal U}_6 &=& \left(\omega_{56}\omega_{46}\omega_{36}\omega_{26}\omega_{16}\right)
\left(\omega_{45}\omega_{35}\omega_{25}\omega_{15}\right)\left(\omega_{34}\omega_{24}\omega_{14}\right)\left(\omega_{23}\omega_{13}\right)\omega_{12}\ . 
\label{eq:U6} 
\end{eqnarray}
Since   $\omega_{45}$  commutes   with  $\omega_{i6}$   for  $i=1,2,3$
[cf.~(\ref{eq:omega1},~\ref{eq:omega2})], we can rewrite~(\ref{eq:U6}) as
\begin{eqnarray}
{\cal U}_6 &=& \left(\omega_{56}\omega_{46}\omega_{45})(\omega_{36}\omega_{26}\omega_{16}\omega_{35}\omega_{25}\omega_{15}\omega_{34}\omega_{24}\omega_{14}\right)\left(\omega_{23}\omega_{13}\omega_{12}\right) \nonumber\\
&= & \left(\begin{array}{cc}
{\bf 1}_3 & {\bf 0}_3\\
{\bf 0}_3 & U_0
\end{array}\right)
\left(\begin{array}{cc}
A & R \\
S & B 
\end{array}\right)
\left(\begin{array}{cc}
V_0 & {\bf 0}_3\\
{\bf 0}_3 & {\bf 1}_3
\end{array}\right) \equiv \left(\begin{array}{cc}
V & R\\ \widetilde{R} & U
\end{array}\right).
\label{eq:U6a}
\end{eqnarray}
Here  the $3\times 3$  unitary matrices  $U_0,V_0$ involve  the mixing
{\em  only}   within  purely  sterile  and   purely  active  neutrinos
respectively,  whereas  the matrices  $R,S$  are  responsible for  the
active-sterile mixing. For convenience, we have defined
\begin{eqnarray}
V\equiv AV_0,~~U\equiv U_0B,~~\widetilde{R}\equiv U_0SV_0.
\end{eqnarray} 
where  $V$ is the  new Pontecorvo-Maki-Nakagawa-Sakata  (PMNS) mixing
matrix for the  active neutrinos. The analytical form  of the matrices
in~(\ref{eq:U6a}), in terms of the  mixing angles and phases, has been
presented in~\cite{xing}.

Finally,  we note  that  the  unitarity of  ${\cal  U}_6$ implies:
\begin{eqnarray}
VV^\dag = AA^\dag = {\bf 1}_3-RR^\dag,\nonumber\\
U^\dag U = B^\dag B = {\bf 1}_3-R^\dag R.
\label{eq:nuty}
\end{eqnarray}
Thus, the matrix $R$ measures the non-unitarity of the PMNS matrix $V$.

%\vfill\eject

\end{document}